\documentclass[prl,twocolumn,showpacs,preprintnumbers,amsmath,%
amssymb,superscriptaddress,floatfix]{revtex4}

\usepackage{graphicx}% Include figure files
\usepackage{dcolumn}% Align table columns on decimal point
\usepackage{bm}% bold math

\begin{document}

\title{Internal mode mechanism for collective energy transport
in extended systems}

\author{Luis Morales--Molina}
\email{Luis.Morales-Molina@uni-bayreuth.de}
\affiliation{Physikalisches Institut, Universit\"at Bayreuth, D-85440 Bayreuth,
Germany}
\affiliation{Grupo Interdisciplinar de Sistemas Complejos
(GISC) and
Departamento de Matem\'aticas,
Universidad Carlos III de Madrid, Avenida de la Universidad 30, 28911
Legan\'es, Madrid, Spain
}%

\author{Niurka R.\ Quintero}
\email{niurka@euler.us.es}
\affiliation{Departamento de F\'\i sica Aplicada I, E.\ U.\ P.,
Universidad de Sevilla, Virgen de \'Africa 7, 41011 Sevilla,
Spain}
\affiliation{Instituto Carlos I de F\'{\i}sica Te\'orica y
Computacional, Universidad de Granada, 18071 Granada, Spain}

\author{Franz G.\ Mertens}
\email{Franz.Mertens@uni-bayreuth.de}
\affiliation{Physikalisches Institut, Universit\"at Bayreuth, D-85440 Bayreuth,
Germany}

\author{Angel S\'anchez}%
\homepage{http://gisc.uc3m.es/~anxo}
\affiliation{%
Grupo Interdisciplinar de Sistemas Complejos (GISC) and
Departamento de Matem\'aticas,
Universidad Carlos III de Madrid, Avenida de la Universidad 30, 28911
Legan\'es, Madrid, Spain
}%

\date{\today}% It is always \today, today,

\begin{abstract}
We study directed energy transport in
homogeneous nonlinear extended systems in the presence of homogeneous
ac forces and dissipation. We show that the
mechanism responsible for unidirectional motion of topological excitations
is the coupling of their internal and translation degrees of freedom. Our results
lead to a selection rule for the existence of such motion based on 
resonances that explains earlier symmetry analysis of this phenomenon.
The direction of motion is found to depend both on the 
initial and the relative phases of the two harmonic drivings,
even in the presence of noise.
\end{abstract}


\maketitle

One intriguing phenomenon that is 
%\cite{Scott},
receiving much attention recently is
net directed motion induced by zero average forces. Originally 
motivated by stochastic models of biomolecular (brownian) motors
\cite{Reimann}, deterministic 
ratchetlike systems \cite{0d,Flach2} are being intensively studied, chiefly 
because of their many potential technological applications \cite{Linke}.
Many such models consist of one or two particles
on a periodic, asymmetric potential and a periodic force (rocking ratchet
\cite{Reimann}). Later, the investigation was generalized to systems
with many interacting particles, from noisy soliton-bearing systems
\cite{Marchesoni,Tsironis} to other 
spatially extended (stochastic and deterministic, overdamped and 
underdamped) systems,  
both theoretically \cite{ext-th} 
and 
from a more applied \cite{ext-app} viewpoint. 

Among this class of problems, 
net transport in {\em homogeneous} extended systems driven by 
{\em homogeneous} ac forces is particularly interesting. 
A paradigmatic example 
is the ac driven, damped sine-Gordon
(sG) equation:
\begin{equation}
\phi_{tt} - \phi_{xx} + \sin(\phi) = -\beta \phi_{t} + f(t).
\label{sG}
\end{equation}
A symmetry analysis, proposed for one-particle
systems in \cite{Flach2} and extended to this problem
\cite{Flach,SZ}, indicated that a directed energy current appeared
if $f(t)$ broke the symmetry $f(t) = - f(t+T/2)$, $T$ being the period
of the external driving. One such choice is 
$f(t)\equiv \epsilon_{1} \sin(\delta t + \delta_{0}) + \epsilon_{2}
\sin(m \delta t + \delta_{0} + \theta)$ (\cite{Flach,SZ} with 
$\delta_0=\pi/2$), a case for which 
numerical simulations of the sG equation confirmed the symmetry analysis
results. 
In what follows, we will refer to $\delta_0$ as the initial phase and
to $\theta$ as the relative phase.
Transport required 
a nonzero topological charge, implying
the existence of sG solitons
(kinks) in the system.
In this respect, we stress that kink-mediated transport is impossible
with only one harmonic for any value of the damping coefficient
$\beta$ \cite{Niurka1}. It was argued in \cite{SZ} that the observed
rectification arises from the nonadiabatic excitation
of internal kink modes and their interaction with the translational
kink motion. This conjecture had no 
rigorous support; rather, it was based on plots of sG soliton 
evolution and on the failure of a collective coordinate (CC) approach
\cite{SIAM} with one degree of freedom, which assumed that
sG solitons behave like rigid particles. An attempt to include the
width degree of freedom has been recently presented in 
\cite{Willis}, where it was concluded that
the directed energy current vanishes unless the width of the kink, $l(t)$,
is a dynamical variable. However, this condition is only a necessary
one: $l(t)$ is a dynamical variable in the one-harmonic case
but the kink velocity is zero for any value of the damping as already
mentioned \cite{Niurka1}. Another point not accounted for in \cite{Willis}
is the connection between the internal mode mechanism and the symmetry 
analysis, which also prohibits motion in other cases
where $l(t)$ is a dynamical variable. Therefore, the reasons for the 
phenomena observed in \cite{Flach,SZ} remained largely obscure. 

In this letter, a different CC approach allows us to
identify the mechanism through which the width oscillation drives the kink
and its relation with the symmetry conditions.
Furthermore, our theory predicts, and numerical simulations of Eq.\ 
(\ref{sG}) confirm, that the direction of motion
depends on the {\em initial phase} of the driving, even
in the presence of additive noise. 
Our CC theory is based on 
an {\em Ansatz}, proposed in \cite{Rice},
for the perturbed kink depending on two CC, $X(t)$
and $l(t)$ (respectively, position and width of 
the kink). It is not difficult to show \cite{Rice,Niurka2,Niurka3}
that the dynamics of these two CC is given by 
\begin{eqnarray} \label{cc-p}
& \, & \frac{dP}{dt}=-\beta P-qf(t), \\
& \, &  \dot{l}^2-2l\ddot{l}-2\beta l\dot{l}= \Omega_{R}^2 l^2\left[1+\frac{P^2}{M_{0}^2}\right]-\frac{1}{\alpha},
\label{cc-l}
\end{eqnarray}
where the momentum $P(t)=M_{0} l_{0} \dot{X}/l(t)$,
$\Omega_{R} = 1/(\sqrt{\alpha} l_{0})$ with $\alpha=\pi^2/12$ is the so-called Rice's frequency,
and 
$M_0=8$, $q=2\pi$  and $l_0=1$ are, respectively, the dimensionless 
kink mass, 
topological charge and unperturbed width.
Equation (\ref{cc-p}) can be solved exactly, and in the 
large time limit ($t\gg \beta^{-1}$) yields 
\begin{equation}\nonumber\label{4}
P(t)=-\sqrt{\epsilon}[a_{1}\sin(\delta t+\delta_{0}-\chi_{1})+a_{2}\sin(m\delta t+\delta_{0}+\theta-\chi_{2})],
\end{equation}
where $\epsilon$ is merely a rescaling parameter in
the perturbation expansion, to be determined later;
     $\chi_{1}=\arctan\left({\delta}/{\beta}\right), \quad \chi_{2}=\arctan\left({m\delta}/{\beta}\right), \quad
        a_{1}={q\epsilon_1}/{\sqrt{\epsilon(\beta^2+\delta^2)}},$ and 
$a_{2}={q\epsilon_2}/{\sqrt{\epsilon(\beta^2+m^2\delta^2)}}.$
%The change of variable $g(t)^2=l(t)$ leads to 
%an Ermakov-type equation $g(t)$,
%given by
%\begin{equation}\label{6}
%\ddot{g}+\beta\dot{g}+\left[\left(\frac{\Omega_{R}}{2}\right)^2+
%\left(\frac{\Omega_{R}}{2M_{0}}\right)^2P^2(t)\right] g =
%\frac{1}{4\alpha g^3},
%\end{equation}
As we are 
interested in the damped ($\beta\neq 0$) case and
Eq.\ (\ref{cc-l}) cannot be solved in that case \cite{Niurka2,Niurka3},
we will study it by 
a perturbative expansion, $l(t)=l_{0}+\epsilon l_{1}(t)+\epsilon^2 l_{2}(t)+\cdots$. 
At order $O(\epsilon)$, we obtain %from Eq.\ (\ref{6})
\begin{equation}\label{10}
\ddot{l}_{1}(t)+\beta \dot{l}_{1}(t)+\Omega_{R}^2 l_{1}(t)
=-{\Omega_{R}^2}P^{2}(t) l_{0}/{2\epsilon M_{0}^{2}}.
%-\frac{\Omega_{R}}{2\epsilon\sqrt{\alpha} M_{0}^{2}}P^{2}(t).
\end{equation}
The key point is 
that, by substituting the expression of $P(t)$
into (\ref{10}), we see that the equation
for $l_{1}(t)$ contains harmonics of frequencies $2 \delta$,
$2 m \delta$ and $(m \pm 1) \delta$, i.e.,
\begin{widetext}
\begin{eqnarray}\label{15}
\nonumber
\ddot{l}_{1}(t)+\beta \dot{l}_{1}(t)+\Omega_{R}^2 l_{1}(t)&=&
%-\frac{\Omega_{R}}{2\epsilon\sqrt{\alpha} M_{0}^{2}}P^{2}(t)=
A_{1}+A_{2}\cos(2\delta t+2\delta_{0}-2\chi_{1})
+A_{3}\cos(2m\delta t+2\delta_{0}+2\theta-2\chi_{2})\\ \nonumber
&+&
A_{4}\cos[(m-1)\delta t +\theta -(\chi_{2}-\chi_{1})]-
A_{4}\cos[(m+1)\delta t+ 2\delta_{0}+ \theta -(\chi_{2}+\chi_{1})],
\end{eqnarray}
where
$A_{1}=-A_{2}-A_{3}$, and
%-\frac{\Omega_{R}}{2\sqrt{\alpha }M_{0}^2}\frac{a_{1}^2
%+a_{2}^2}{2},\>\>\>
$A_{2}={\Omega_{R}a_{1}^2}/{4\sqrt{\alpha }M_{0}^2},\>
A_{3}={\Omega_{R}a_{2}^2}/{4\sqrt{\alpha }M_{0}^2},\>
A_{4}=-{\Omega_{R}a_{1}a_{2}}/{2\sqrt{\alpha }M_{0}^2}.$
After transients elapse, we find
\begin{eqnarray} \label{16}
l_{1}(t)&=&\frac{A_{1}}{\Omega_{R}^2}+
\frac{A_{2}\sin(2\delta t+2\delta_{0}-2\chi_{1}+\tilde{\theta}_{2})}
{\sqrt{(\Omega_{R}^2-4\delta^2)^2+4\beta^2\delta^2}}+
\frac{A_{3}\sin(2m\delta t+2\delta_{0}+2\theta-2\chi_{2}+
\tilde{\theta}_{2m})}{\sqrt{(\Omega_{R}^2-4m^2\delta^2)^2+
4m^2\beta^2\delta^2}}\\ \nonumber
&&+\frac{A_{4}\sin[(m-1)\delta t +\theta -(\chi_{2}-\chi_{1})+
\tilde{\theta}_{m-1}]}{\sqrt{(\Omega_{R}^2-(m-1)^2\delta^2)^2+
\beta^2(m-1)^2\delta^2}}
%\\
%&&
-\frac{A_{4}\sin[(m+1)\delta t + 2\delta_{0}+
\theta -(\chi_{2}+\chi_{1})+\tilde{\theta}_{m+1}]}
{\sqrt{(\Omega_{R}^2-(m+1)^2\delta^2)^2+\beta^2(m+1)^2\delta^2}}, 
\end{eqnarray}
\end{widetext}
where
$\tilde{\theta}_{m}=
\arctan\left[\left({\Omega_{R}^2-m^2\delta^2}\right)/{m\beta\delta}\right].$
A cumbersome but otherwise 
trivial calculation yields the harmonics 
contained in $l_{2}(t)$,
collected in Table \ref{tab1}. 

Next, we need to
compute the average velocity over one period $T=2 \pi/\delta$:
In the CC approach, we
use the definition of the momentum and find
\begin{equation}\label{1}
\langle\dot{X}(t)\rangle=\frac{1}{T}\int_{0}^{T}\frac{P(t)l(t)}{M_{0}l_{0}}dt.
\end{equation}
%Considering the expansion  (\ref{7}), this expression
%can be written as
%\begin{eqnarray}\label{13}
%\nonumber
%\langle\dot{X}(t)\rangle&=&\frac{1}{T}\int_{0}^{T}\frac{P(t)(l_{0}+
%\epsilon l_{1}(t)+\epsilon^2 l_{2}(t)+...)}{M_{0}l_{0}}dt\\
%&=&\langle\dot{X}_{0}(t)\rangle+\epsilon \langle\dot{X}_{1}(t)\rangle+\epsilon^2 \langle\dot{X}_{2}(t)\rangle+...
%\end{eqnarray}
%Therefore the mean velocity can be  calculated analytically in some approach
At $O(\epsilon^{0})$, the averages $\langle P(t)\rangle$
and $\langle\dot{X}_{0}(t)\rangle$ vanish trivially; therefore, 
net kink motion can only arise
in next order.  
By straightforward calculations from
Eqs.\ (\ref{16}) and (\ref{1}) we find for $m=2$ that, for large enough times,
\begin{widetext}
\begin{eqnarray}
\epsilon\langle\dot{X}_{1}\rangle=
\frac{q^3\Omega_{R}^2\epsilon_{1}^2\epsilon_{2}}{8M_{0}^{3}
(\beta^2+\delta^2)\sqrt{\beta^2+4\delta^2}}\left(\frac{2\cos[\delta_{0}-
\theta+(\chi_{2}-2\chi_{1})-\tilde{\theta}_{1}]}{\sqrt{(\Omega_{R}^2-
\delta^2)^2+\beta^2\delta^2}} \right. \label{17}
%\\ \nonumber
\left.
-\frac{\cos[\delta_{0}-\theta+(\chi_{2}-2\chi_{1})+
\tilde{\theta}_{2}]}{\sqrt{(\Omega_{R}^2-4\delta^2)^2+
4\beta^2\delta^2}}\right).
\end{eqnarray}
\end{widetext}
{}From
Eq. (7) we see that for $\epsilon$ to be small
the prefactor on the rhs has to be much smaller than 1.
A definite, verifiable prediction from this asymptotic expression
is the existence of a {\em nonzero velocity for $m=2$}, with 
a sinusoidal dependence on $\delta_{0}$ and $\theta$.
This means
that {\em the velocity depends on both the initial and the relative phases;} 
indeed, by letting $\delta_0\equiv\delta t_0$ in Eq.\ (\ref{sG}) 
and changing variables 
to $t'=t+t_0$, it can be
immediately seen that an initial phase $\delta_0$ is equivalent to a
relative phase $\theta'=\theta-(m-1)\delta_0$ for a kink with its center
shifted to $x_0+Vt_0$. The dependence of the velocity on
$\theta$ agrees with  (and explains) 
\cite{SZ,Flach}, whereas the dependence on $\delta_0$ is a totally new
result.%, in principle contradictory to what is reported in \cite{SZ}. 
Nevertheless,
these analytical results as well as the 
numerical simulations we present below strongly support the present 
conclusion. 

\begin{table}
\caption{Harmonic content of the first contributions to the
perturbative expansion of $l(t)$.}\label{tab1}
\begin{center}
\begin{tabular}{|c|c|c|}
\hline
 harmonic & $l_{1}$ & $l_{2}$ \\
\hline
$m$ & $2\delta$, $2m\delta$, $(m\pm 1)\delta$ & $2\delta$, $4\delta$, $2m\delta$, 
$4m\delta$, $(m\pm 1)\delta$, \\
\, & \, & $2(m\pm 1)\delta$, $(m\pm 3)\delta$, $(3m\pm 1)\delta$\\
\hline
2 & $\delta$, $2\delta$, $3\delta$, $4\delta$ & $\delta$, $2\delta$, $3\delta$, $4\delta$, $5\delta$, $6\delta$, $7\delta$, $8\delta$\\
\hline
3 & $2\delta$, $4\delta$, $6\delta$  &  $2\delta$, $4\delta$, $6\delta$, $8\delta$, $10\delta$,  $12\delta$ \\
%\hline
%4 & $2\delta$, $3\delta$, $5\delta$, $8\delta$ &  $\delta$,  $2\delta$, $3\delta$,  $4\delta$, $5\delta$, $7\delta$,\\
%\, & \, & $9\delta$, $11\delta$, $13\delta$, $16\delta$\\
\hline
\end{tabular}
\end{center}
\end{table}
For the case $m=3$, the average velocity is zero
at all orders, 
a result confirmed by direct numerical simulation of the
full sG equation (\ref{sG}) as we will see below. 
The reason can be understood by looking at Table \ref{tab1}: 
For $m=3$, the frequencies of the ac force
(or the momentum) are odd harmonics ($\delta$ and $3\delta$), whereas
the width of the kink oscillates only with even harmonics
($2n\delta$, $n \in \mathbb{N}$). This leads us to our main conclusion, namely
the mechanism for the appearance of net motion and the corresponding
selection rules. Equations (\ref{cc-p}) and (\ref{cc-l}) show that 
the force acts on the kink width through $P^2(t)$, whereas $P(t)$
itself is in turn 
inversely proportional to $l(t)$. This coupling is the responsible 
for the net kink motion, but for it to be actually possible, 
{\em the harmonic content of the effective 
force $P^2(t)$ acting on the width
degree of freedom must be able to resonate with it.}
This is evident from Eq.\ (\ref{1}), in which the integral is nonzero
only if $l(t)$ contains at least one of the harmonics of $P(t)$.
It is important to realize that this condition is much more restrictive
than that found in \cite{Willis}, where only the necessity of $l(t)$
being a dynamic variable was pointed out. We have just seen that this 
is indeed necessary, but that additional, crucial resonance conditions 
have to be fulfilled. 
Interestingly, our theory shows also that dissipation can change or
even revert the kink velocity, see Eq.\ (\ref{17}), in agreement with
the numerical results in \cite{Flach,SZ}. A more detailed discussion 
of this point is forthcoming \cite{future}. 
%\noindent
\begin{figure}
\vspace*{2mm}
\includegraphics[width=6.0cm,height=3.0cm]{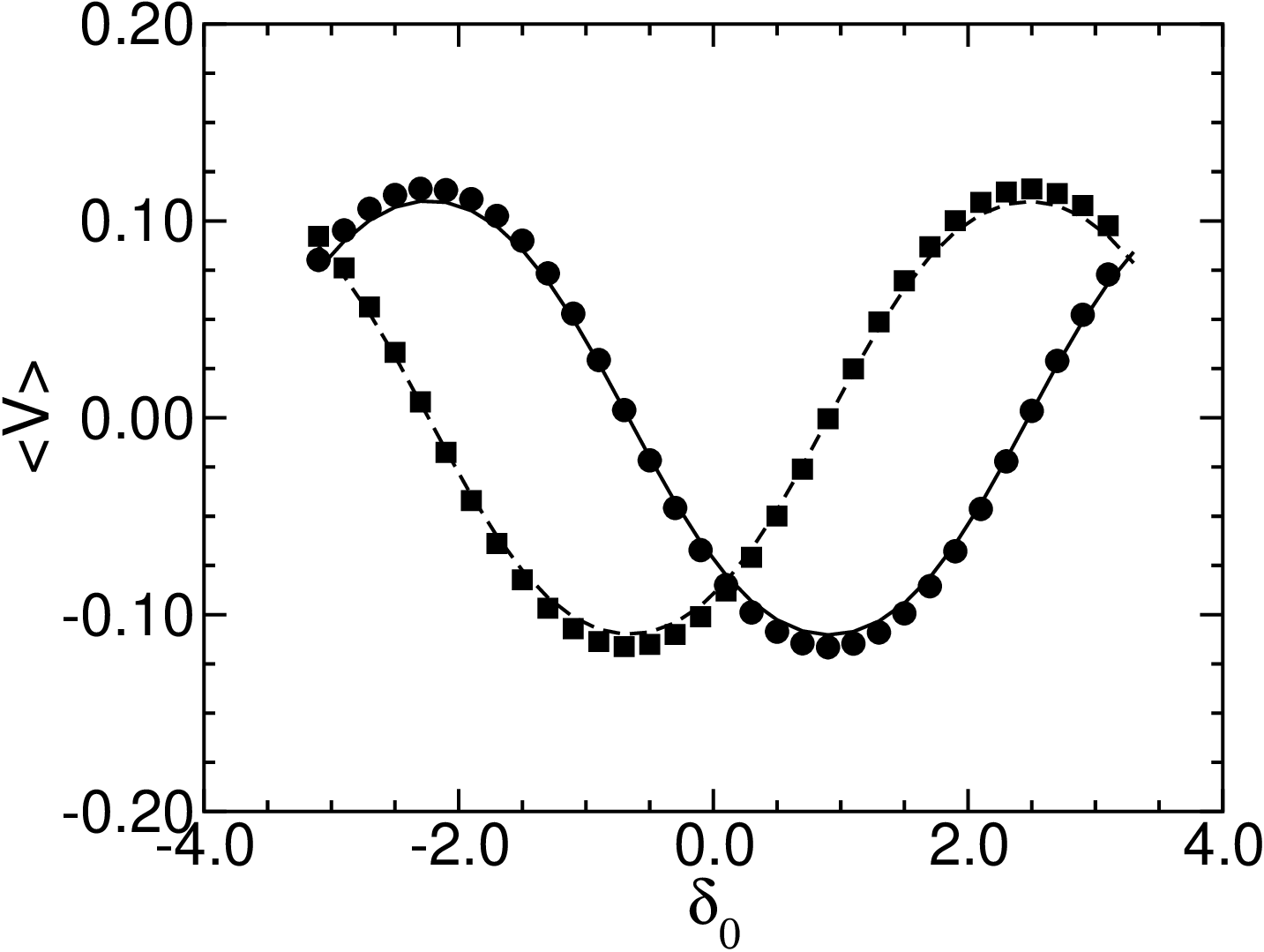}
\caption{\label{fig:faseinicial} 
Dependence of the kink velocity on the initial phase. Parameters are 
$\epsilon_1=\epsilon_2=0.2$, $\beta=0.05$, $\delta=0.1$. Relative phase $\theta=\pi/2$: 
solid line, CC theory; filled circles, simulation results.
Relative phase $\theta=0$:
dashed line, CC theory; squares, simulation results.
}
\end{figure}

\begin{figure}
\vspace*{2mm}
\includegraphics[width=6.0cm,height=3.0cm]{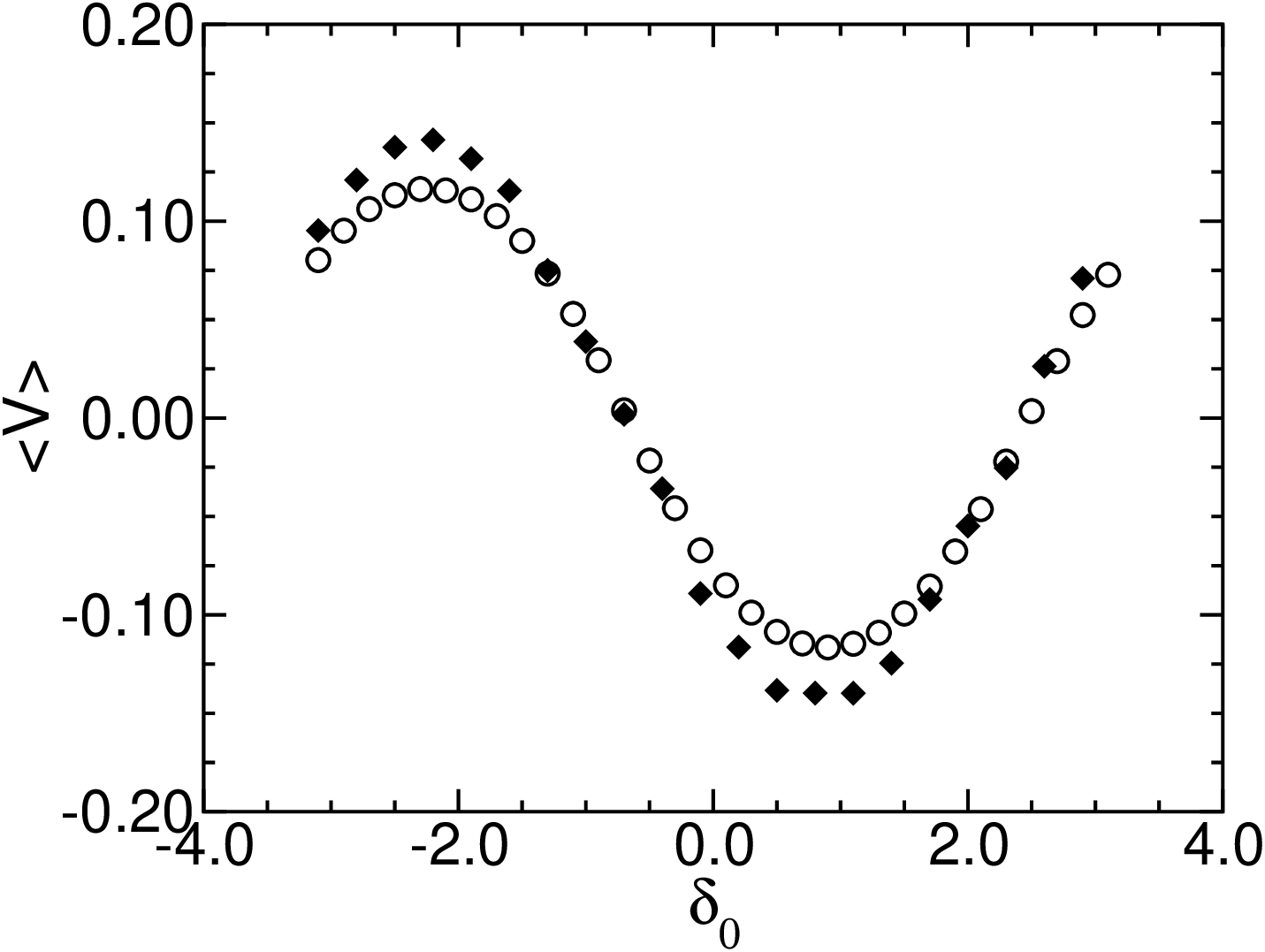}
\caption{\label{fig:ruido} 
Dependence of the kink velocity on the initial phase for 
relative phase $\theta=\pi/2$ in the deterministic ($D=0$, 
empty circles) and the stochastic ($D=0.03$, diamonds) cases.
%Line is only a guide to the eye.
Other parameters are as in Fig.\ \ref{fig:faseinicial}.}
\end{figure}

\begin{figure}
\vspace*{2mm}
\includegraphics[width=6.0cm,height=3.0cm]{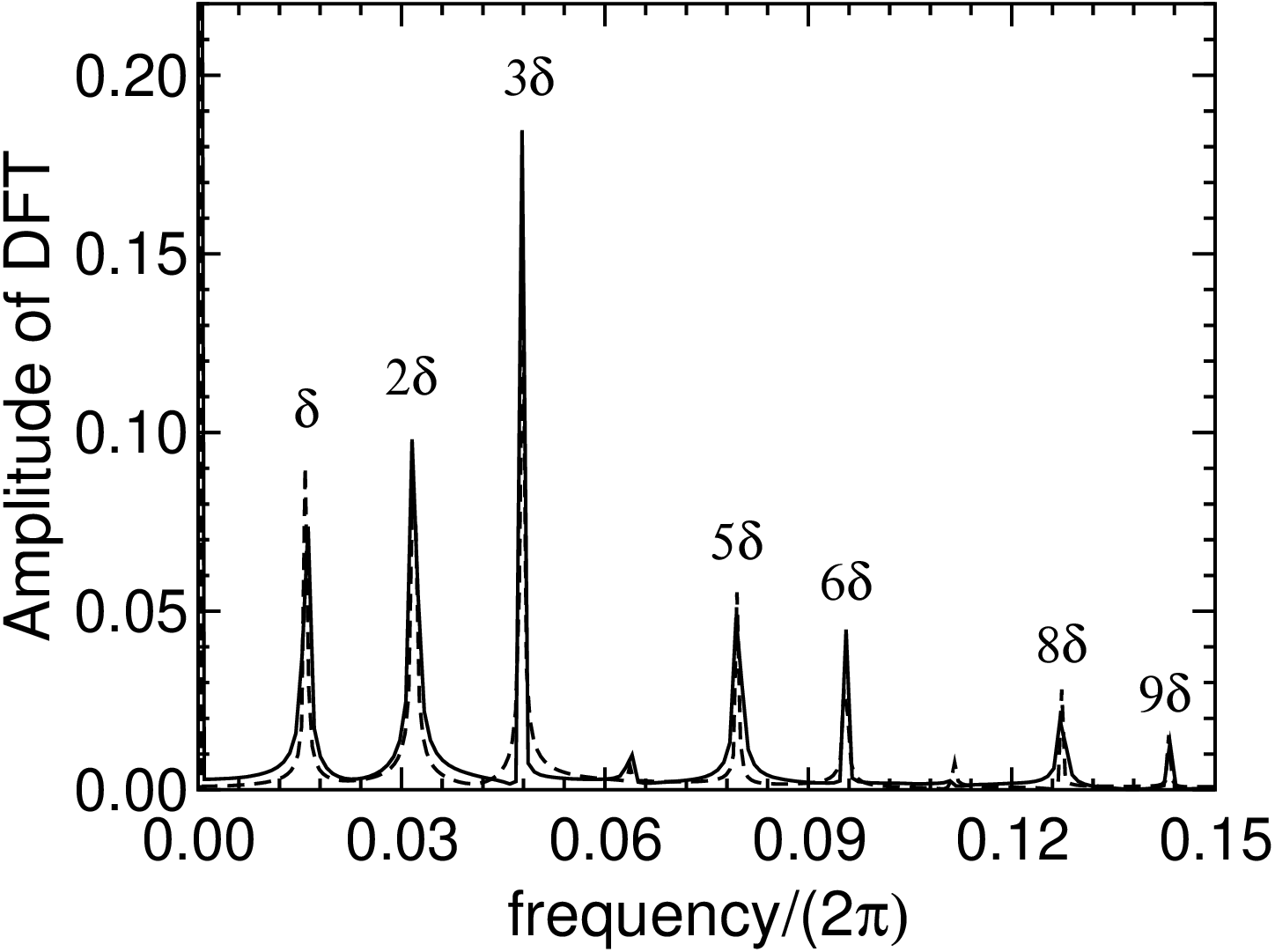}\\[1mm]
\includegraphics[width=6.0cm,height=3.0cm]{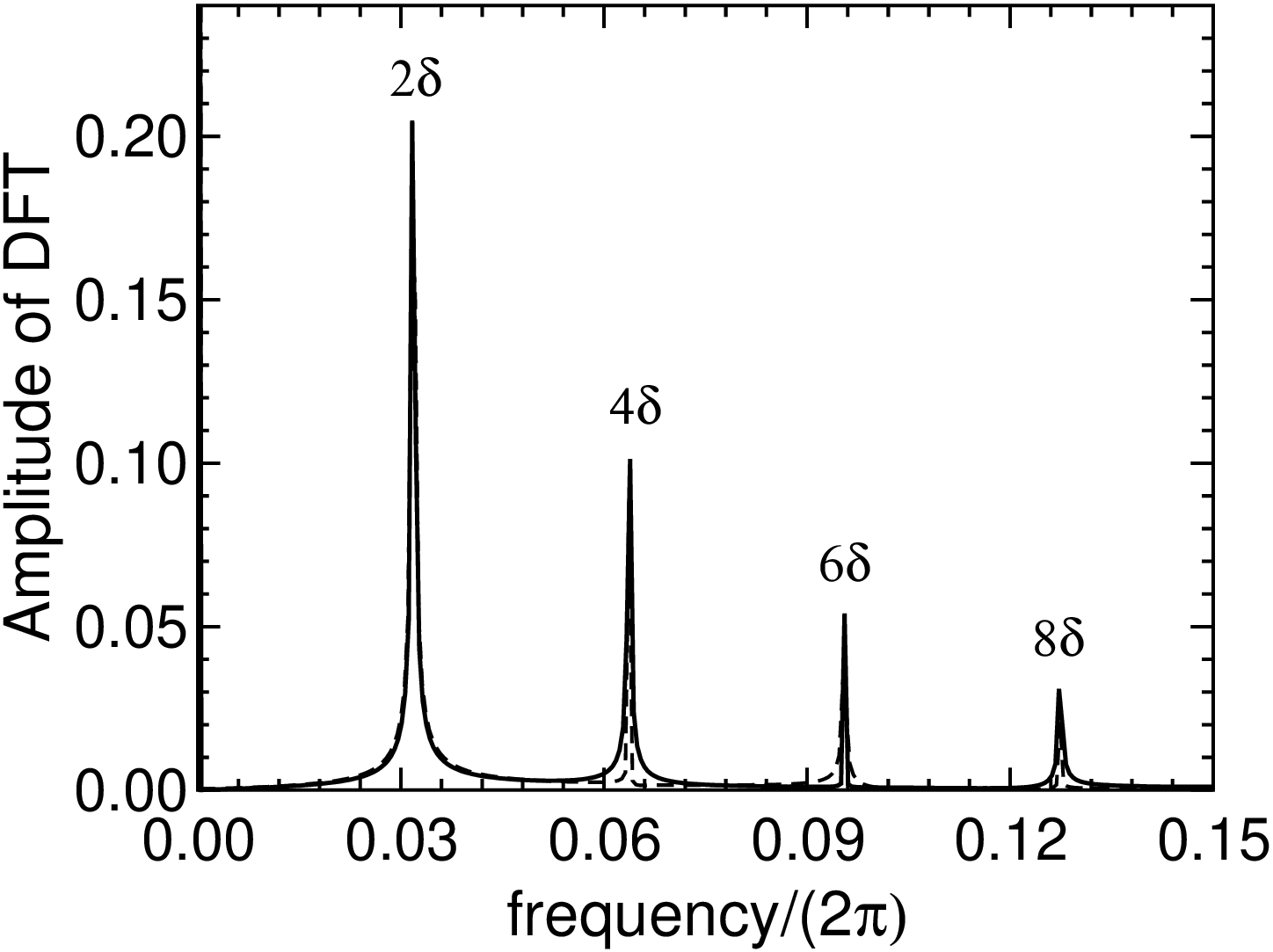}%\\[1mm]
\caption{\label{fig:dft} 
Discrete Fourier Transform of the kink width.
Upper panel: $m=2$; lower panel: $m=3$.
Solid line: amplitude measured in simulations. Dashed line: 
numerical integration of the CC equations.
Parameters are as in Fig.\ \ref{fig:faseinicial} 
for relative phase $\theta=\pi/2$ and $\delta_0=-2.5$.}
\end{figure}

These predictions from the CC
approximation must be confirmed by 
numerical solution of the full partial differential
equation (\ref{sG}). This we do by using the Strauss-V\'azquez
scheme \cite{vaz}, on systems of length
of $L=100, 1000$, with steps $\Delta t=0.01$,  $\Delta x=0.1$,
free boundary conditions, and a kink at rest as initial condition.
Instead of the perturbative expressions
(which are
only qualitatively correct unless $\epsilon << 1$),
to assess the validity of 
our theory we numerically integrate
Eq.\ (\ref{cc-l}) and the equation for the velocity obtained from 
the expression of $P(t)$, Eq.\ (\ref{cc-p}), 
with a fourth-order Runge-Kutta method.
Our main results are shown in Figs.\ 
\ref{fig:faseinicial}--\ref{fig:dft}; they fully confirm the accuracy, 
even quantitative, of our approach. 
Figure \ref{fig:faseinicial} exhibits clearly the
sinusoidal dependence of 
the velocity as a function of the initial phase. The dependence on 
$\theta$ is also seen as a simple shift when changing from $\theta=0$
to $\theta=\pi/2$. The agreement with the CC results is perfect. 
As a further check of the robustness of this dependence, 
following \cite{Flach} we have simulated Eq.\ (\ref{sG}) with an additional
additive gaussian white noise term with variance $D$.
While one could in principle think that this 
noise would suppress the initial phase dependence, Fig.\ \ref{fig:ruido} shows
that the opposite is the case: the {\em noise enhances the dependence on the
initial phase}, increasing the maximum values of the velocity while keeping
the same general sinusoidal dependence and the location of the zeros. It
is tempting to conclude from this plot that the noise, at least if it is 
not very large ($D<<1$),
assists the process of energy transfer between the width
and the translation degrees of freedom, activating it. Finally, Fig.\ 
\ref{fig:dft} makes it clear that our main result, namely the
interpretation of the physics of 
the problem, is indeed true, by showing the harmonic content of $l(t)$ 
for $m=2$ and $3$. In this case, the agreement between our
CC theory and the full numerical simulation of Eq.\ (\ref{sG}) is 
indeed impressive, and validates firmly our resonance criterion for
net kink motion.
It is important to stress that the present theory does not apply to 
the net motion found for $m=3$ in 
\cite{SZ}. We have confirmed their result in our simulations, which 
allowed us to realize that
this is an altogether different phenomenon: first, it appears only above 
a (moderately large, $\epsilon_i\gtrsim 0.4$) threshold amplitude, and 
second, it is 
induced by the {\em 
kink wings}, which are highly distorted in the process yielding the
CC picture inappropriate (even kink-antikink pairs are created).
%\noindent

In conclusion, we 
have found that the symmetry
conditions set forth in \cite{Flach,SZ} have their physical origin in the
mechanism of the directed motion: the indirect action of the force
through the coupling of the translational and width degrees of freedom. 
To make net motion possible, 
this indirect driving has to resonate with
the available frequencies for the width. This interpretation does not
contradict the non-existence of internal modes in 
sG kinks, shown in  
\cite{Niurka3}, because 
external forces can induce%
, via excitation of certain phonons, 
behavior similar to the one expected from an intrinsic internal 
mode \cite{Niurka5,Gonzalez}. The fact that
a force with only one harmonic would not drive the damped 
sG kink \cite{Niurka1},
and that 
two harmonics are needed to simultaneously excite the width oscillations 
and induce net motion fits nicely in this picture. On the other hand, 
this point rises the question as to the generality of our results, in 
view of the fact that most kink-bearing systems do have internal modes. 
To answer this question, 
we have studied the same problem in the framework of 
the $\phi^4$ model, reaching the same conclusions 
\cite{future}):
Indeed, the intrinsic internal mode of $\phi^4$ kinks makes the phenomenon
even more noticeable, making us confident on the wide applicability of
this work.

Another important conclusion is 
the dependence of the velocity on the initial 
phase $\delta_0$, not mentioned in earlier work
\cite{Flach,SZ}. We note that this dependence allows much more 
flexibility in controlling the kink velocity, providing an alternative
to the use of the relative phase suggested earlier.
%Note that, strictly speaking, this implies that this system does not 
%behave in a ratchet-like fashion, as an average over initial conditions 
%would yield zero velocity.
On the other hand, this 
may have important consequences for applications 
as a way of separating, e.g., fluxons in long Josephson junctions
\cite{ext-app}. Interestingly, such superconducting devices provide the 
best possible laboratory to verify our results. 
This experimental confirmation is crucial in order
to ascertain their applicability. Given the accuracy with which the sG
equation describes long Josephson junctions, and the fact that an external
force like the one proposed in this and earlier works \cite{Flach,SZ} is 
easy to implement, we hope that the corresponding measurements will soon
be carried out. A conclusive, positive verification of our theory would
yield the picture we provide here very useful 
in that and related contexts. 
%\noindent

\begin{acknowledgments}
This work has been
supported by the Ministerio de Ciencia y Tecnolog\'\i a of Spain 
through grants BFM2001-3878-C02 (NRQ), 
BFM2000-0006 and BFM2003-07749-C05-01 
(AS), by the Junta de Andaluc\'{\i}a under the
project FQM-0207, by DAAD (Germany)
A0231253/Ref.\ 314, and by the International Research Training Group
`Nonequilibrium Phenomena and Phase Transitions in Complex Systems'.
%(DFG, Germany).
\end{acknowledgments}

\end{document}